\documentclass[12pt,a4paper]{article}
\usepackage{amssymb}

\usepackage[left=5em]{geometry}
\usepackage{amsmath}
\usepackage{graphicx}

\title{Weak Gravity Conjecture and Holographic Dark Energy Model with Interaction and Spatial Curvature}
\author{\textbf{Cheng-Yi Sun\footnote{cysun@mails.gucas.ac.cn; ddscy@163.com}\ $^{,a}$\
}\\ \\
 {$^a$\small Institute of Modern Physics, Northwest University,}\\
     \small Xian 710069, P.R. China.}

\begin{document}
\maketitle
\begin{abstract}
In the paper, we apply the weak gravity conjecture to the
holographic quintessence model of dark energy. Three different
holographic dark energy models are considered: without the
interaction in the non-flat universe; with interaction in the flat
universe; with interaction in the non-flat universe. We find that
only in the models with the spatial curvature and interaction term
proportional to the energy density of matter, it is possible for the
weak gravity conjecture to be satisfied. And it seems that the weak
gravity conjecture favors an open universe and the decaying of
matter into dark energy.
\end{abstract}



\section{Introduction}
Increasing evidence suggests that the expansion of our universe is
being accelerated \cite{Supernova,WMAP,LSS}. Within the framework of
the general relativity, the acceleration can be phenomenally
attributed to the existence of a mysterious exotic component with
negative pressure, namely the dark energy \cite{dark energy1,dark
energy2}, which dominates the present evolution of the universe.
However, we know little about the nature of dark energy. The dark
energy problem has become one of the focuses in the fields of both
cosmology and theoretical physics today.

The most nature, simple and important candidate for dark energy is
the Einstein's cosmological constant, which can fit the observations
well so far. But the cosmological constant is plagued with the
well-known fine-tuning and cosmic coincidence problems \cite{dark
energy1,dark energy2}. Another candidate for dark energy is
scalar-field dark energy model. So far, a wide variety of
scalar-filed dark energy models have been proposed, such as
quintessence \cite{quitessence}, phantom \cite{phantom}, $k$-essence
\cite{kessence}, tachkyon \cite{tachyon}, quintom \cite{quintom},
hessence \cite{hessence}, etc. Usually, the scalar-field models are
regarded as an low-energy effective description of the underlying
theory of dark energy. Other dynamical dark energy models include
Chaplygin gas models \cite{Chaplygin}, braneworld models
\cite{braneworld}, etc.

A lot of efforts have been made to solve the dark energy problem,
but no effort seems to be successful so far. Actually, it is
generally believed that the dark energy problem is in essence an
issue of quantum gravity. However a complete theory of quantum
gravity is still unknown. Then it becomes natural for physicists to
explore the nature of dark energy just in light of some fundamental
principles of quantum gravity. The holographic principle is commonly
believed to be such a principle \cite{holographicP}. Based on the
principle, the holographic dark energy models has been suggested
\cite{HDE}. The model has been studied widely, and supported by
various observations (see citations of Ref.\cite{HDE}). And even, it
is found that the holographic dark energy model is favored by the
anthropic principle \cite{0410095}.

On the other hand, it is generally believed that string theory is
the most promising theory of quantum gravity. Recent progress
\cite{kklt} suggests that there exist a vast number of
semi-classical consistent vacua in string theory, named
\emph{Landscape} \cite{landscape}. However, not all semi-classical
consistent vacua are actually consistent on the quantum level, and
these actually inconsistent vacua are called \emph{Swampland}
\cite{vafa}. Self-consistent landscape is surrounded by the
swampland. The weak gravity conjecture (WGC) is suggested to be a
new criterion to distinguish the landscape from the swampland
\cite{wgc1,wgc2}. The conjecture can be most simply stated as
gravity is the weakest force. For a four-dimensional U(1) gauge
theory, WGC implies that there is an intrinsic UV cutoff \cite{wgc1}
\[
  \Lambda\leq gM_p,
\]
where $g$ is the gauge coupling constant and $M_p$ is the Planck
scale. Furthermore, in \cite{0703071}, it is argued that WGC also
indicates an intrinsic UV cutoff for the scalar field theories with
gravity, e.g.
\[
  \Lambda\leq \lambda^{1/2} M_p
\]
for $\lambda\phi^4$ theory. In the slow-roll inflation model with
the potential $V(\phi)\sim\lambda\phi^4$, Hubble constant can be
taken as the IR cutoff for the field theory. Then the requirement
that the IR cutoff should be lower than the UV cutoff indicates
\cite{0703071}
\begin{equation}
  \label{HLmabda}\frac{\lambda^{1/2}\phi^2}{M_p}\sim H\leq\Lambda\leq \lambda^{1/2}
  M_p, \quad\texttt{or,}\quad  \phi\leq M_p
\end{equation}
This leads the author in \cite{0706.2215} to conjecture that the
variation of the inflaton during the period of inflation should be
less than $M_p$,
\begin{equation}
  \label{deltaPhiMp}|\Delta\phi|\leq M_p.
\end{equation}
And it is found that this can make stringent constraint on the
spectral index of the inflation model \cite{0706.2215}.

Recently, the criterion (\ref{deltaPhiMp}) has been used in
\cite{0708.2760} to explore the quintessence model of dark energy,
in \cite{0710.1406} to study Chaplygin gas models \cite{Chaplygin},
and in \cite{1005.2466} to study the agegraphic dark energy model
\cite{0708.0884}. Then if the holographic dark energy scenario is
assumed to be the underlying theory of dark energy, and the
low-energy scalar field can be used to describe it effectively
\cite{HQDE}, can the weak gravity conjecture, i.e.,
$\Delta\phi<M_p$, be satisfied? In the direction, some work has been
done \cite{0709.1517,0806.2415}. It is found that the holographic
quintessence model does not satisfied the conjecture
\cite{0806.2415}. However, in \cite{0709.1517,0806.2415}, only the
non-interacting holographic quintessence in the flat universe is
discussed. In \cite{0910.3855}, it is found that when simultaneously
considering the interaction and spatial curvature in the holographic
dark energy model, the parameter space is amplified much more. Then
it may be possible for the weak gravity conjecture to be satisfied
in the interacting holographic quintessence model with spatial
curvature. Here we will discuss the problem.

In the paper, we will first recall the interacting holographic dark
energy model in the non-flat universe. Then we will discuss the
possible theoretical constraints on the holographic quintessence
model from the weak gravity conjecture and try to find the
possibility for the conjecture to be satisfied within the parameter
space displayed in \cite{0910.3855}. Finally,  conclusion will be
given.

\section{Holographic Dark Energy Model with Interaction and Space Curvature}

The Friedmann-Robertson-Walker (FRW) universe is described by the
line element
\begin{equation}
  \label{FRWmetric}
  ds^2=-dt^2+a^2(t)\Big(\frac{dr^2}{1-kr^2}+r^2d\Omega^2\Big),
\end{equation}
where $a(t)$ is the scale factor, and $k$ is the curvature parameter
with $k=-1,0,1$ corresponding to a spatially open, flat and closed
universe, respectively. The Friedmann equation is
\begin{equation}
  \label{FE}
  3M_p^2\Big(H^2+\frac{k}{a^2}\Big)=\rho_m+\rho_D,
\end{equation}
where $M_p=(8\pi G)^{-1/2}$, $\rho_m$ is the energy density of
matter and $\rho_D$ is the energy density of dark energy. By
defining
\begin{equation}
  \label{fractionalED}
  \Omega_k=\frac{\rho_k}{\rho_c}=\frac{k}{H^2a^2},\quad \Omega_D=\frac{\rho_D}{\rho_c},
  \quad \Omega_m=\frac{\rho_m}{\rho_c},
\end{equation}
where $\rho_k=k/a^2$ and $\rho_c=3M_p^2H^2$, we can rewrite the
Friedmann equation as
\begin{equation}
  \label{reFD}
  1+\Omega_k=\Omega_D+\Omega_m.
\end{equation}
In the holographic dark energy model, the energy density $\rho_D$ is
assumed to be \cite{0403127}
\begin{equation}
  \label{HDE}
  \rho_D=3d^{2}M_p^2R_h^{-2},
\end{equation}
where $d$ is a constant parameter, $R_h=ar(t)$ and $r(t)$ satisfies
\cite{0404229}
\begin{equation}
  \label{rt}
  \int^{r(t)}_0{\frac{dr}{\sqrt{1-kr^2}}}=\int_t^{+\infty}{\frac{dt}{a(t)}}.
\end{equation}
For a closed universe, $k=1$, from the equation above we have
\begin{equation}
  \label{arcsingR}
  \arcsin{(\sqrt{k}r)}=\sqrt{k}\int_t^{+\infty}{\frac{dt}{a(t)}}.
\end{equation}
Together with Eq.(\ref{HDE}), we have
\begin{equation}
  \label{r}
  \arcsin{\Big(\sqrt{k\frac{3d^2M_p^2}{a^2\rho_D}}\Big)}=\sqrt{k}\int_t^{+\infty}{\frac{dt}{a(t)}}.
\end{equation}
The derivative of the equation with respect to $t$ gives
\begin{equation}
  \label{dRhodDt}
  \dot{\rho}_D=-3H\rho_D\times\frac{2}{3}\Big(1-\sqrt{\frac{\Omega_D}{d^2}-\Omega_k}\Big).
\end{equation}
The equation is obtained by using $k=1$. But it can be easily
checked that Eq.(\ref{dRhodDt}) holds in the cases of $k=0$ and
$k=-1$, too. We can define an effective equation of state parameter
$w^{\text{eff}}_D$ by
\begin{equation}
  \label{DefWeff}
  \dot{\rho}_D+3H(1+w^{\text{eff}}_D)\rho_D=0.
\end{equation}
Then comparing Eqs.(\ref{dRhodDt}) and (\ref{DefWeff}), we have
\begin{equation}
  \label{weff}
  w^{\text{eff}}_D=-\frac{1}{3}\Big(1+2\sqrt{\frac{\Omega_D}{d^2}-\Omega_k}\Big).
\end{equation}
Using Eqs.(\ref{fractionalED}), we may rewrite Eq.(\ref{dRhodDt}) as
\cite{0910.3855}
\begin{equation}
  \label{OmegaH1}
  \sqrt{\frac{\Omega_DH^2}{d^2}-\frac{k}{a^2}}=\frac{\dot{\Omega}_D}{2\Omega_D}+H+\frac{\dot{H}}{H}.
\end{equation}

Now consider some interaction between holographic dark energy and
matter \cite{0910.3855}
\begin{align}
  \label{matterCL}
  \dot{\rho}_m+3H\rho_m&=Q,\\
  \label{darkEnergyCL}
  \dot{\rho}_D+3H(1+w_D)\rho_D&=-Q,
\end{align}
where $Q$ denotes the phenomenological interaction term. In
Ref.\cite{0910.3855} three types of interaction are considered,
\begin{align}
  \label{Q1}
  Q_1&=-3bH\rho_D\\
  \label{Q2}
  Q_2&=-3bH(\rho_D+\rho_m)\\
  \label{Q3}
  Q_3&=-3bH\rho_m.
\end{align}
Following Ref.\cite{0910.3855}, for convenience, we uniformly
express the interaction term as $Q_i =-3bH\rho_c\Omega_i$, where
$\Omega_i =\Omega_D,1+\Omega_k$ and $\Omega_m$, for $i=1,2$ and $3$,
respectively.

From Eqs.(\ref{FE}), (\ref{matterCL}) and (\ref{darkEnergyCL}), we
can obtain
\begin{equation}
  \label{wP}
  w_D=-\frac{1}{3H\rho_D}\Big(2\frac{\dot{H}}{H}\rho_c-2H\rho_k\Big)-\rho_c-\rho_k.
\end{equation}
Substituting the equation into Eq.(\ref{darkEnergyCL}) and using
Eq.(\ref{fractionalED}), we can have \cite{0910.3855}
\begin{equation}
  \label{OmegaH2}
  2(\Omega_D-1)\frac{\dot{H}}{H}+\dot{\Omega}_D+H(3\Omega_D-3-\Omega_k)=3bH\Omega_i,
\end{equation}
where $i=1,2$ and $3$. From Eqs.(\ref{OmegaH1}) and (\ref{OmegaH2}),
finally we get the following two equations governing the evolution
of the interacting holographic dark energy in the non-flat universe
\begin{align}
  \label{dHdz}
  \frac{d\widetilde{H}}{dz}&=-\frac{\widetilde{H}}{1+z}\Omega_D\Bigg(\frac{3\Omega_D-\frac{\Omega_{k0}(1+z)^2}{\widetilde{H}^2}-3-3b\Omega_i}{2\Omega_D}-1
                                                                    +\sqrt{\frac{\Omega_D}{d^2}-\frac{\Omega_{k0}(1+z)^2}{\widetilde{H}^2}}\Bigg),\\
  \label{dOmegaDdz}
  \frac{d\Omega_D}{dz}&=-\frac{2\Omega_D(1-\Omega_D)}{1+z}\Bigg(\sqrt{\frac{\Omega_D}{d^2}-\frac{\Omega_{k0}(1+z)^2}{\widetilde{H}^2}}-1
                                                             -\frac{3\Omega_D-\frac{\Omega_{k0}(1+z)^2}{\widetilde{H}^2}-3-3b\Omega_i}{2(1-\Omega_D)}\Bigg),
\end{align}
where $1+z=\frac{1}{a}$, $\widetilde{H}\equiv\frac{H}{H_0}$ and
hereafter the subscript $0$ denotes the present value of the
corresponding parameter. And we have used
$\Omega_k=\frac{\Omega_{k0}(1+z)^2}{\widetilde{H}^2}$ and $a_0=1$.

\section{Holographic Quintessence Model and Weak Gravity Conjecture}

For a single-scalar-field quintessence model with potential
$V(\phi)$, the energy density and pressure of the quintessence
scalar field are
\begin{align}
  \label{phiED}
  \rho_\phi&=\frac{1}{2}\dot{\phi}^2+V(\phi),\\
  \label{phiP}
  p_\phi&=\frac{1}{2}\dot{\phi}^2-V(\phi).
\end{align}
So the equation of state is
\begin{equation}
  \label{eos}
  w_\phi=\frac{\rho_\phi}{p_\phi}=\frac{\frac{1}{2}\dot{\phi}^2+V}{\frac{1}{2}\dot{\phi}^2-V}
\end{equation}
From Eqs.(\ref{phiED}) and (\ref{eos}), we can obtain easily
\begin{equation}
  \label{rhoWphi}
  \rho_\phi=\frac{\dot{\phi}^2}{1+w_\phi}
\end{equation}
Without loss of generality, we may assume $dV/d\phi<0$ and
$\dot{\phi}>0$. Thus from Eq.(\ref{rhoWphi}), we may have
\begin{equation}
  \label{dPhidt}
  \dot{\phi}=\sqrt{(1+w_\phi)\rho_\phi}
\end{equation}
Then the weak gravity conjecture tells us
\cite{0703071,0708.2760,0709.1517,0806.2415}
\begin{equation}
  \label{wgc}
  \begin{split}
    1\geq\frac{|\Delta\phi(z)|}{M_p}&=\int{\frac{\dot{\phi}}{M_p}dt}\\
                                    &=\int^z_0{\sqrt{3[1+w_\phi(z')]\Omega_\phi(z')}\frac{dz'}{1+z'}},
  \end{split}
\end{equation}
where $\Omega_\phi=\rho_\phi/\rho_c$.

In the holographic quintessence model, the holographic dark energy
is assumed to be described by the effective scalar field. Then
naturally we have
\begin{equation}
  \label{phiEDHDE}
  \rho_\phi=\rho_D\Rightarrow\Omega_\phi=\Omega_D
\end{equation}
Here following Ref.\cite{0509040}, we identify $w_\phi$ with
$w^{\text{eff}}_D$, instead of $w_D$,
\begin{equation}
  \label{wphiweff}
  w_\phi=w^{\text{eff}}_D=-\frac{1}{3}\Big(1+2\sqrt{\frac{\Omega_D}{d^2}-\Omega_k}\Big).
\end{equation}
We must require $d\geq1$ so that $w_\phi \geq -1$. Substituting
Eq.(\ref{wphiweff}) into Eq.(\ref{wgc}), then the weak gravity
conjecture for the holographic quintessence with interaction and
spatial curvature reads
\begin{equation}
  \label{wgc0}
  \begin{split}
    1\geq\frac{|\Delta\phi(z)|}{M_p}=\int^z_0{\sqrt{2\Big(1-\sqrt{\frac{\Omega_D(z')}{d^2}
                                                           -\frac{\Omega_{k0}(1+z)^2}{\widetilde{H}^2}}\Big)\Omega_D(z')}\frac{dz'}{1+z'}}.
  \end{split}
\end{equation}
where $\Omega_D$ and $\widetilde{H}$ can obtained by numerically
solving Eqs.(\ref{dHdz}) and (\ref{dOmegaDdz}) if the initial values
$\Omega_{m0}$ and $\Omega_{k0}$, and the values of the constant
parameters $b$ and $d$ are given.

The case of $\Omega_{k0}=0$ and $b=0$ has been discussed in
Ref.\cite{0709.1517,0806.2415}, and it is found the weak gravity
conjecture cannot be satisfied for a holographic quintessence model.
Here we expect the weak gravity conjecture may be satisfied when
both the interaction and spatial curvature are considered.

Since
\begin{equation}
  \label{OmegaDd}\Omega_D=\frac{d^2}{H^2R_h^2},
\end{equation}
then larger $b$ indicates bigger $\Omega_D$ and more difficult for
Eq.(\ref{wgc0}) to be satisfied. On the other hand, we must require
$d\geq1$ so that $w_\phi>-1$. So in the paper, we will focus on the
case of $d=1$. Then Eq.(\ref{wgc0}) becomes
\begin{equation}
  \label{wgcd1}
  \begin{split}
    1\geq\frac{|\Delta\phi(z)|}{M_p}=\int^z_0{\sqrt{2\Big(1-\sqrt{\Omega_D(z')
                                                           -\frac{\Omega_{k0}(1+z)^2}{\widetilde{H}^2}}\Big)\Omega_D(z')}\frac{dz'}{1+z'}}.
  \end{split}
\end{equation}

\subsection{the non-interacting holographic quintessence model with spatial curvature}

With $b=0$, we can rewrite Eq.(\ref{wgcd1}) as
\begin{equation}
  \label{wgcd1b0}
  \begin{split}
    1\geq\frac{|\Delta\phi(z)|}{M_p}=\int^z_0{\sqrt{2\Big(1-\frac{\Omega_{m0}(1+z)^\alpha}{\widetilde{H}^2}\Big)\Omega_D(z')}\frac{dz'}{1+z'}},
  \end{split}
\end{equation}
where $\alpha=3$, and we have used
$\Omega_m=\frac{\Omega_{m0}(1+z)^\alpha}{\widetilde{H}^2}$ and
Eq.(\ref{reFD}). Then the equation above tells us that larger
$\Omega_{m0}$ will make it  easier for Eq.(\ref{wgcd1b0}) to be
satisfied. However, in Ref.\cite{0806.2415}, it is shown that
Eq.(\ref{wgcd1b0}) can not be satisfied in the flat universe even
for $\Omega_{m0}=0.35$.

\begin{figure}
\centering
\renewcommand{\figurename}{Fig.}
\includegraphics[scale=0.6]{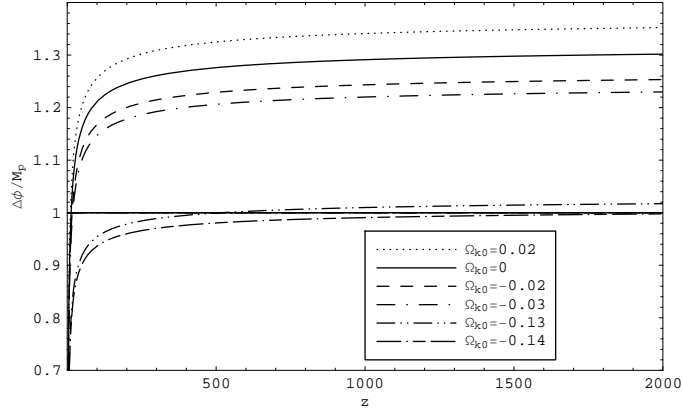}
\caption{$\Delta\phi(z)/M_p$ versus the redshift $z$ in the
non-interacting holographic quintessence model with spatial
curvature for fixed $\Omega_{m0}=0.34$ and different
$\Omega_{k0}$.\label{KHDE}}
\end{figure}

Now let us consider the effect of the spatial curvature. From
Eq.(\ref{reFD}), we know that larger $\Omega_k$ indicates larger
$\Omega_D$ and then makes it more difficult for Eq.(\ref{wgcd1b0})
to be satisfied. So negative $\Omega_k$ will make it easier for
Eq.(\ref{wgcd1b0}) to be satisfied than that in the flat universe.
We plot the result of $\Delta\phi/M_p$ versus the redshift $z$ in
Fig.\ref{KHDE}. In the case, we find that it is impossible for the
weak gravity conjecture to be satisfied within the parameter space
displayed in FIG.4 in \cite{0910.3855}. Actually, in order to match
the weak gravity conjecture, we should take
$\Omega_{k0}\lesssim-0.14$ which has been far outside the range of
$\Omega_{k0}$ displayed in FIG. 4 in \cite{0910.3855}. In
Fig.\ref{KHDE}, we have fixed $\Omega_{m0}=0.34$ which is slightly
bigger than the maximum value of $\Omega_{m0}$ displayed in FIG.4 in
Ref.\cite{0910.3855}, since smaller $\Omega_{m0}$ would make it more
difficult for Eq.(\ref{wgcd1b0}) to be satisfied.

\subsection{the interacting holographic quintessence model without spatial curvature}

With the interaction between the dark energy and matter, we can
rewrite Eq.(\ref{wgcd1}) as
\begin{equation}
  \label{wgcd1k0}
  \begin{split}
    1\geq\frac{|\Delta\phi(z)|}{M_p}=\int^z_0{\sqrt{2\Big(1-\frac{\Omega_{m0}(1+z)^{\alpha(z)}}{\widetilde{H}^2}\Big)\Omega_D(z')}\frac{dz'}{1+z'}},
  \end{split}
\end{equation}
where $\alpha(z)$ is defined by
\begin{equation}
  \label{alpha}
  \ln{\frac{\rho_m}{\rho_{m0}}}=\alpha(z)\ln(1+z).
\end{equation}
Here the Friedmann equation reads
\[
  1=\Omega_m+\Omega_D.
\]
Since the form of Eq.(\ref{wgcd1k0}) is similar to that of
Eq.(\ref{wgcd1b0}), we can get a similar conclusion that larger
$\Omega_{m0}$ will make it easier for Eq.(\ref{wgcd1k0}) to be
satisfied.

For the three types of interaction, we can uniformly express the
conservation law Eq.(\ref{matterCL}) as
\begin{equation}
  \label{matterCLi}
  \dot{\rho}_m=-3H\Big(1+b\frac{\Omega_i}{\Omega_m}\Big)\rho_m,
\end{equation}
where $i=1,2$ and $3$. From Eqs.(\ref{alpha}) and (\ref{matterCLi}),
we have
\begin{equation}
  \label{alphab}
  \alpha(z)=\frac{3}{\ln(1+z)}\int_0^z{\Big(1+b\frac{\Omega_i(z')}{\Omega_m(z')}\Big)\frac{dz'}{1+z'}}.
\end{equation}
Obviously, $\alpha=3$ for $b=0$, $\alpha<3$ for $b<0$ and $\alpha>3$
for $b>0$. Or, in the other words, smaller $b$ indicates smaller
$\alpha$, and then will make it more difficult for
Eq.(\ref{wgcd1k0}) to be satisfied.

\begin{figure}
\centering
\renewcommand{\figurename}{Fig.}
\includegraphics[scale=0.6]{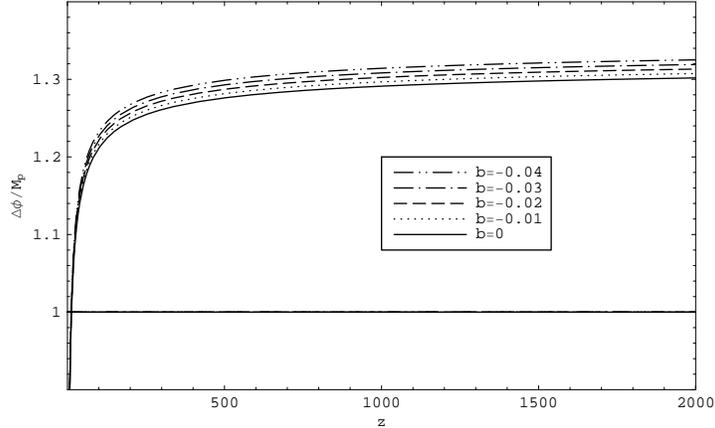}
\caption{$\Delta\phi(z)/M_p$ versus the redshift $z$ in the
holographic quintessence model with the interaction term $Q=Q_1$ in
the flat universe for fixed $\Omega_{m0}=0.34$ and different
$b$.\label{IHDE1}}
\end{figure}

\begin{figure}
\centering
\renewcommand{\figurename}{Fig.}
\includegraphics[scale=0.6]{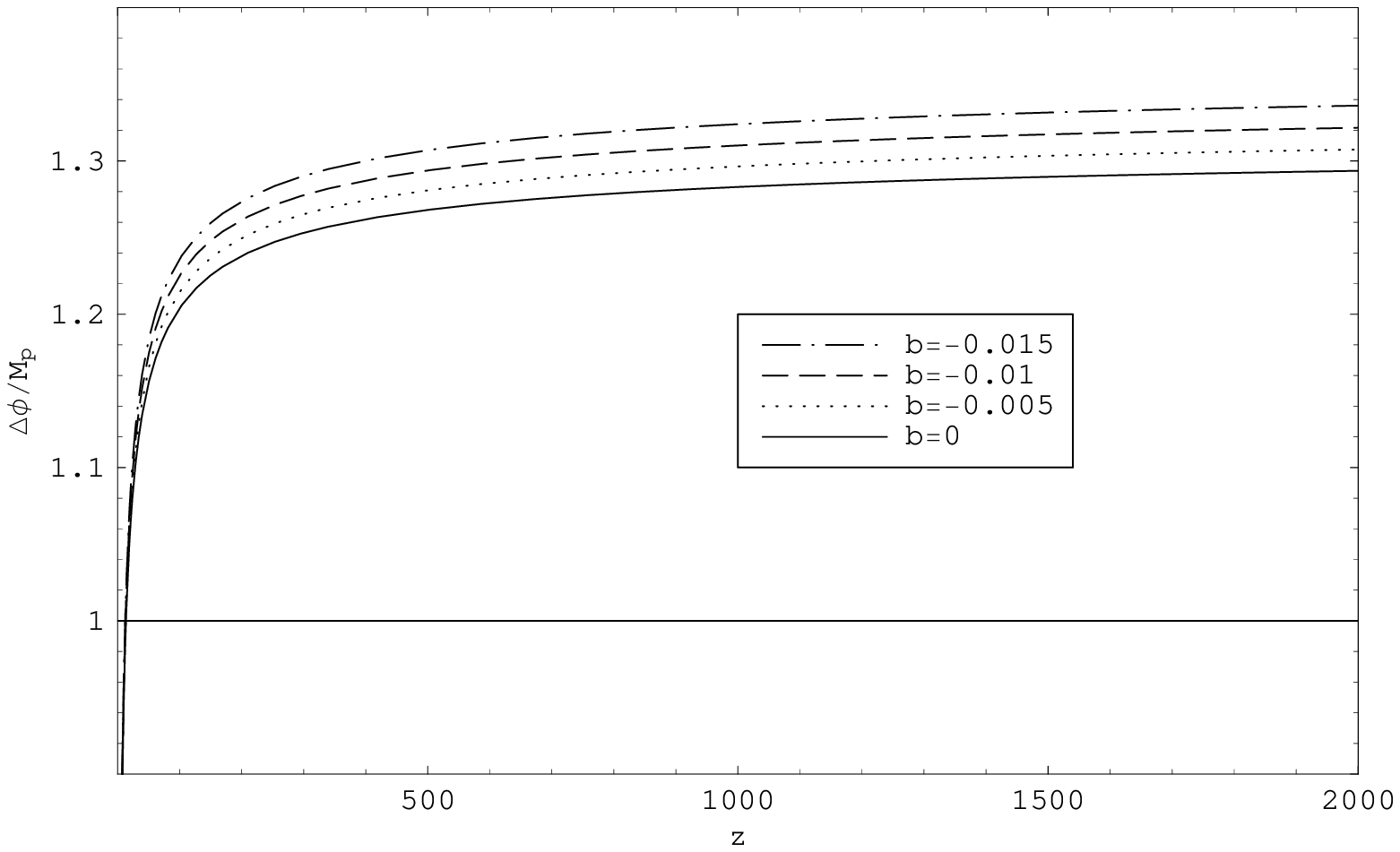}
\caption{$\Delta\phi(z)/M_p$ versus the redshift $z$ in the
holographic quintessence model with the interaction term $Q=Q_2$ in
the flat universe for fixed $\Omega_{m0}=0.34$ and different
$b$.\label{IHDE2}}
\end{figure}

On the other hand, it has been shown in \cite{0806.2415} that the
weak gravity conjecture cannot be satisfied in the holographic
quintessence models in the flat universe with $b=0$ even for
$\Omega_{m0}=0.35$. Since smaller $b$ or $\Omega_{m0}$ would make it
more difficult for the weak gravity conjecture to be satisfied, and
$\Omega_{m0}=0.35$ has been far beyond the range of $\Omega_{m0}$
displayed in FIG.1 in Ref.\cite{0910.3855},  we can conclude that
the weak gravity conjecture cannot be satisfied in the flat universe
with non-positive $b$. Our conclusion is illustrated by the results
displayed in Fig.\ref{IHDE1} and Fig.\ref{IHDE2}.

\begin{figure}
\centering
\renewcommand{\figurename}{Fig.}
\includegraphics[scale=0.7]{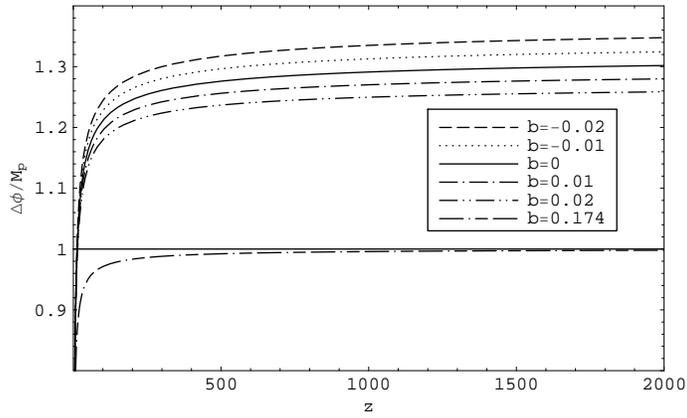}
\caption{$\Delta\phi(z)/M_p$ versus the redshift $z$ in the
holographic quintessence model with the interaction term $Q=Q_3$ in
the flat universe for fixed $\Omega_{m0}=0.34$ and different
$b$.\label{IHDE3}}
\end{figure}

For the models with the interaction term $Q=Q_1$ or $Q=Q_2$, only
the case of $b\leq0$ is regarded as the realistic physical
situation. The reason is that in the two types of models, positive
$b$ will lead $\rho_m$ to become negative in the far future. Then we
know that, in the holographic quintessence models with interaction
term $Q=Q_1$ or $Q=Q_2$ in the flat universe, the weak gravity
conjecture can not be satisfied, as shown in Fig.\ref{IHDE1} and
Fig.\ref{IHDE2}.

If $Q=Q_3$, the conservation law of matter is
\begin{equation}
  \label{matterCL3}
  \dot{\rho}_m=-3H(1+b)\rho_m\Rightarrow \rho_m\propto a^{-3(1+b)}.
\end{equation}
Then $b>0$ is also in the realistic physical region since $\rho_m$
will never become negative in the case. But we find even for
positive $b$, in the range of $b$ given in FIG.2 in
Ref.\cite{0910.3855}, Eq.(\ref{wgcd1k0}) can not be satisfied yet.
Our results are shown in Fig.\ref{IHDE3}. Naively, in order to match
the weak gravity conjecture, we should take $b\gtrsim 0.174$ which
is far beyond the range of $b$ given in FIG.2 in
Ref.\cite{0910.3855}.

In the subsection, we still fix $\Omega_{m0}=0.34$ which is slightly
bigger than the maximum value of $\Omega_{m0}$ displayed in FIG.2 in
Ref.\cite{0910.3855}, since smaller $\Omega_{m0}$ would make it more
difficult for Eq.(\ref{wgcd1k0}) to be satisfied.

\subsection{the holographic quintessence model with interaction and spatial curvature}

When both the interaction and spatial curvature are considered, we
still have Eqs.(\ref{wgcd1k0}) and (\ref{alphab}). Of course, now
the Friedmann equation includes the spatial curvature:
\[
  1+\Omega_k=\Omega_m+\Omega_D.
\]
The analysis of the effects of $\Omega_{m0},\Omega_{k0}$ and $b$
still works in the models with interaction and spatial curvature:
larger $\Omega_{m0}$, smaller $\Omega_{k0}$ or larger $b$ will make
it easier for the weak gravity conjecture to be satisfied. From
FIG.5 in \cite{0910.3855}, we know $b$ and $\Omega_{k0}$ is
anti-correlated: larger $b$ corresponds to smaller $\Omega_{k0}$.
Then there might exit the combinations of $\Omega_{m0}$,
$\Omega_{k0}$ and $b$ that satisfies the weak gravity conjecture
Eq.(\ref{wgcd1k0}).

\begin{figure}
\centering
\renewcommand{\figurename}{Fig.}
\includegraphics[scale=0.55]{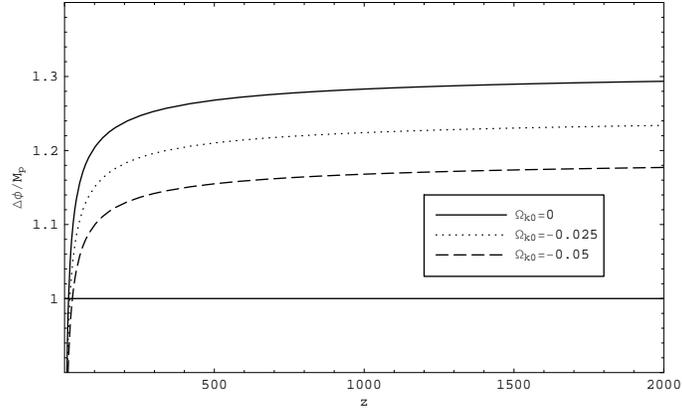}
\caption{$\Delta\phi(z)/M_p$ versus the redshift $z$ in the
holographic quintessence model with the interaction term $Q=Q_1$ and
spatial curvature for $\Omega_{m0}=0.35$, $b=0$ and different
$\Omega_{k0}$.\label{KIHDE1}}
\end{figure}

\begin{figure}
\centering
\renewcommand{\figurename}{Fig.}
\includegraphics[scale=0.55]{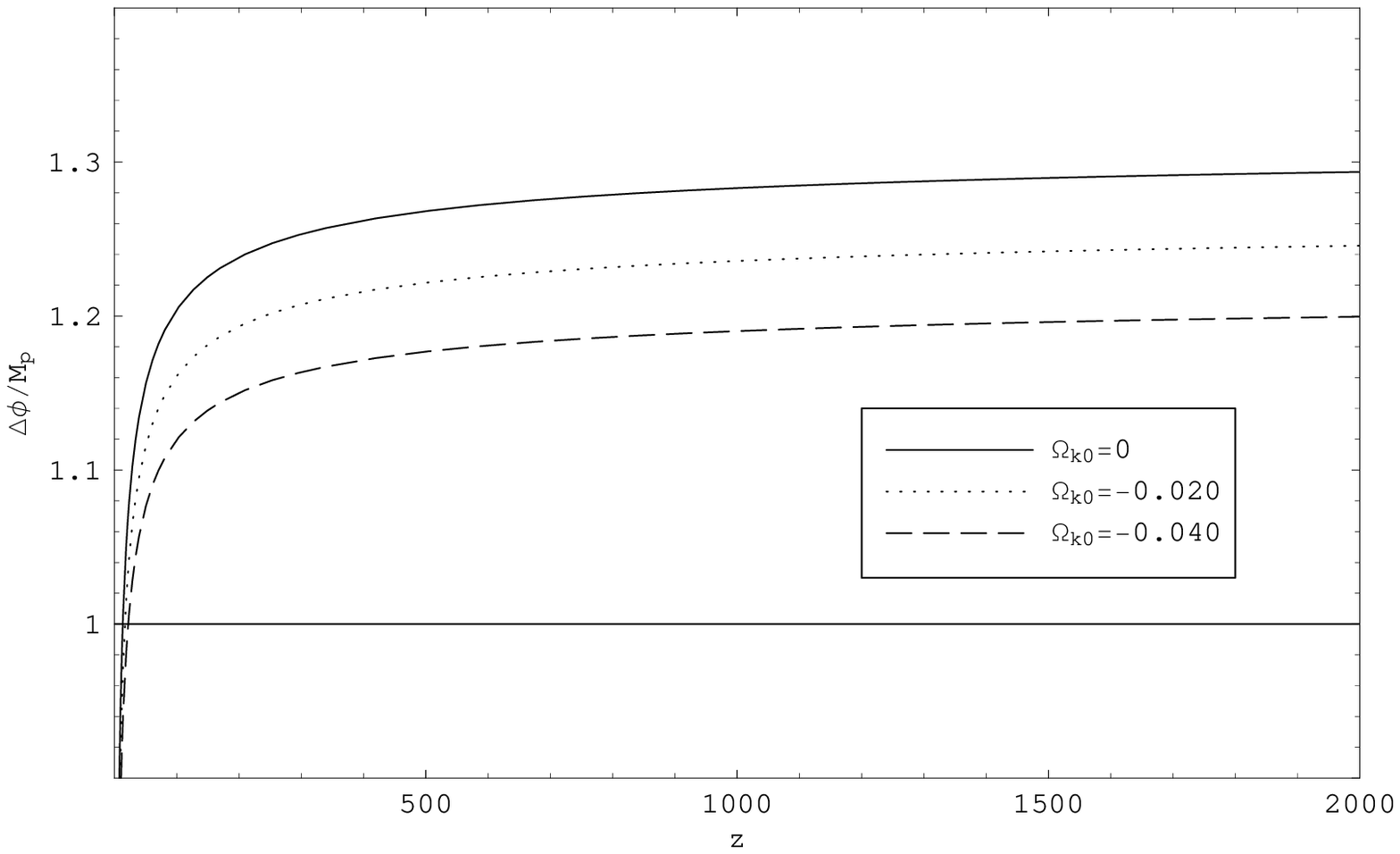}
\caption{$\Delta\phi(z)/M_p$ versus the redshift $z$ in the
holographic quintessence model with the interaction term $Q=Q_2$ and
spatial curvature for $\Omega_{m0}=0.35$, $b=0$ and different
$\Omega_{k0}$. \label{KIHDE2}}
\end{figure}

However, unfortunately, for the models with the interaction term
$Q=Q_1$ or $Q=Q_2$ in the non-flat universe, we can not find any
combination of $\Omega_{m0}$, $\Omega_{k0}$ and $b$ within the
parameter space given in FIG.5 in Ref.\cite{0910.3855} so that
Eq.(\ref{wgcd1k0}) is satisfied. We display our results in
Fig.\ref{KIHDE1} and Fig.\ref{KIHDE2}. In the two figures, we fix
$\Omega_{m0}=0.35$ and $b=0$, since smaller $\Omega_{m0}$ or
negative $b$ would make it more difficult for Eq.(\ref{wgcd1k0}) to
be satisfied, and larger $\Omega_{m0}$ or positive $b$ are not
physical.

\begin{figure}
\centering
\renewcommand{\figurename}{Fig.}
\includegraphics[scale=0.55]{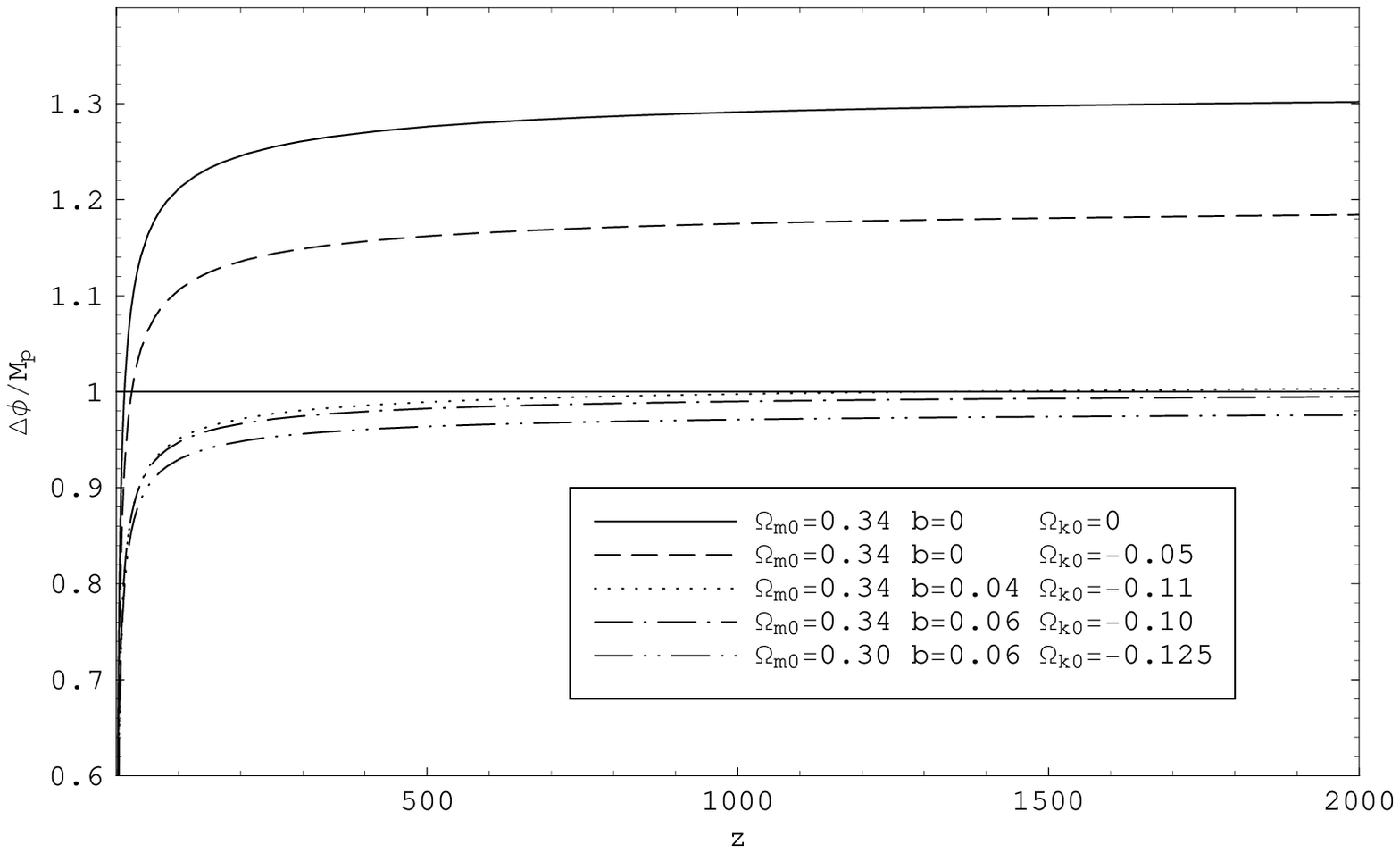}
\caption{$\Delta\phi(z)/M_p$ versus the redshift $z$ in the
holographic quintessence model with the interaction term $Q=Q_1$ and
spatial curvature for different $\Omega_{m0}$, $b$ and
$\Omega_{k0}$.\label{KIHDE3}}
\end{figure}

For the models with the interaction term $Q=Q_3$, We find that the
weak gravity conjecture can not be satisfied within the parameter
space displayed in FIG.5 in \cite{0910.3855} if $b=0$. However, if
$b\gtrsim 0.04$ and $\Omega_{k0}\lesssim -0.10$, the combinations of
$b$ and $\Omega_{k0}$ within the parameter space in FIG.5 of
Ref.\cite{0910.3855} which satisfy Eq.(\ref{wgcd1k0}) can be found.
Our results are shown in Fig.\ref{KIHDE3}.

\section{Conclusion}

In the paper, we have discussed the theoretical limits on the
holographic quintessence model of dark energy from the weak gravity
conjecture. Since the non-interacting holographic quintessence model
without spatial curvature has been investigated in
\cite{0709.1517,0806.2415}, here we consider the other three cases
separately: without interaction in the non-flat universe; with
interaction in the flat universe; with interaction in the non-flat
universe. Here, we use the observational constraints given in
\cite{0910.3855}. We find that the the weak gravity conjecture can
not be satisfied even in the the holographic quintessence models
with interaction in the flat universe or the models without
interaction in the non-flat universe. In \cite{0910.3855}, it is
shown that the parameter space is amplified when simultaneously
considering the interaction and spatial curvature. Then we might
expect that it should be possible for the weak gravity conjecture to
be satisfied in the models with interaction and spatial curvature.

However, we find that the models with the interaction term $Q=Q_1$
or $Q=Q_2$ in the non-flat universe are still inconsistent with the
weak gravity conjecture within the parameter space in
\cite{0910.3855}. Fortunately, we find that, in the models with the
spatial curvature and interaction term $Q=Q_3$, it is possible for
the weak gravity conjecture to be satisfied. A roughly necessary
condition for the weak gravity conjecture to be satisfied is shown:
$b\gtrsim 0.04$ and $\Omega_{k0}\lesssim -0.10$.

Then our results indicate that only the holographic dark energy
models with the  spatial curvature and interaction term $Q=Q_3$ may
be described by a consistent low-energy effective scalar field
theory. And it seems that the weak gravity conjecture favors an open
universe and the decaying of matter into dark energy.

\section*{Acknowledgments}
This work is supported by the Natural Science Foundation of the
Northwest University of China under Grant No. NS0927.

\end{document}